\documentclass{elsart}
\usepackage{amssymb}
\usepackage[dvips]{graphicx}

\makeatletter
\newdimen\z@ \z@=0pt 
\newskip\z@skip \z@skip=0pt plus0pt minus0pt
\def\m@th{\mathsurround=\z@}
\def\ialign{\everycr{}\tabskip\z@skip\halign} 
\def\eqalign#1{\null\,\vcenter{\openup\jot\m@th
  \ialign{\strut\hfil$\displaystyle{##}$&$\displaystyle{{}##}$\hfil
      \crcr#1\crcr}}\,}
\makeatother

\newcommand{\aff}[2]{Dipartimento di Fisica dell'Universit\`a #1 e Sezione INFN, #2, Italy.}
\newcommand{\affd}[1]{Dipartimento di Fisica dell'Universit\`a e Sezione INFN, #1, Italy.}

\begin{document}
\begin{flushright}
    KLOE Note n$^o$ 188

    July 2003
\end{flushright}
\begin{frontmatter}

\title{Searching for $\eta \to \gamma \gamma \gamma$ decay}

\vskip -1cm

\collab{The KLOE Collaboration}

\author[Na] {A.~Aloisio},
\author[Na]{F.~Ambrosino},
\author[Frascati]{A.~Antonelli},
\author[Frascati]{M.~Antonelli},
\author[Roma3]{C.~Bacci},
\author[Frascati]{G.~Bencivenni},
\author[Frascati]{S.~Bertolucci},
\author[Roma1]{C.~Bini},
\author[Frascati]{C.~Bloise},
\author[Roma1]{V.~Bocci}
% ...

\clearpage
\address[Bari]{\affd{Bari}}
\address[Frascati]{Laboratori Nazionali di Frascati dell'INFN, 
Frascati, Italy.}
\address[Karlsruhe]{Institut f\"ur Experimentelle Kernphysik, 
Universit\"at Karlsruhe, Germany.}
\address[Lecce]{\affd{Lecce}}
\address[Na]{Dipartimento di Scienze Fisiche dell'Universit\`a 
``Federico II'' e Sezione INFN,
Napoli, Italy}
\address[Pisa]{\affd{Pisa}}
\address[Energ]{Dipartimento di Energetica dell'Universit\`a 
``La Sapienza'', Roma, Italy.}
\address[Roma1]{\aff{``La Sapienza''}{Roma}}
\address[Roma2]{\aff{``Tor Vergata''}{Roma}}
\address[Roma3]{\aff{``Roma Tre''}{Roma}}
\address[StonyBrook]{Physics Department, State University of New 
York at Stony Brook, USA.}
\address[Trieste]{\affd{Trieste}}
\address[Virginia]{Physics Department, University of Virginia, USA.}
\address[Beijing]{Permanent address: Institute of High Energy 
Physics, CAS,  Beijing, China.}
\address[Moscow]{Permanent address: Institute for Theoretical 
and Experimental Physics, Moscow, Russia.}
\address[Tbilisi]{Permanent address: High Energy Physics Institute, Tbilisi
  State University, Tbilisi, Georgia.}

\begin{abstract}
   Data collected by the KLOE experiment in years 2001/2002 for a total integrated luminosity of 410 pb$^{-1}$,
corresponding to 17 million of produced $\eta$'s from the $\Phi \to \eta
\gamma$ radiative decay, have been analyzed to look for the C-violating decay 
$\eta\to\gamma\gamma\gamma$. 
    The signal is  searched in 
events with 4 photons in the final state by using  the spectrum
of the most energetic photon.  
    The background is evaluated from the data themself, doing a fifth degree
 polynomial fit outside the signal region, and extrapolating inside it. 
    No signal has been observed and taking into account a selection
 efficiency \makebox{$\epsilon = 20.3 \%$} we get from a likelihood fit
 Br$(\eta\to\gamma\gamma\gamma
) \le 1.6\times10^{-5}$ at 95\% CL. The procedure has a systematic
ambiguity at the 35 $\%$ level.
\end{abstract}
\end{frontmatter}

\section{Introduction}

At the Frascati $\phi$ factory DA$\Phi$NE \cite{Dafne} a large number of $\eta$ 
mesons is produced via the radiative decay of the $\phi$ meson.   \\

The $\phi$ mesons are produced in the collision of $e^{+}e^{-}$ beams into the DA$\Phi$NE
collider and 1.298 \% of them decay into the $\eta \gamma$ final state. Up to now $17\times
10^{6}$ $\eta$ mesons have been
 produced and are present in 410 pb$^{-1}$  collected 
 by the KLOE experiment  in years 2001-2002.\\

The $\eta$ meson is an even eigenstate of the charge coniugation operator \emph{C},
while a photon is an odd \emph{C} eigenstate. So any decay of the $\eta$ meson
into a final state with an odd number of photons violates the \emph{C} symmetry.
In the Standard Model, the C  symmetry is exactly conserved in both strong and electromagnetic decays, but it is violated in weak decays due to the V-A structure of
the weak couplings. In the framework of the Standard Model the decay rate of the $\pi ^0 \rightarrow 3 \gamma$ has been evaluated \cite{Dicus}, and generalizing this result to the $\eta$ case \cite{Nefkens&Price}, one obtains Br$(\eta \rightarrow 3 \gamma) < 10 ^{-12}$. \\
For this reason the discovery of a larger decay rate would be
a clear signal of SM violation. At the moment all the predictions of alternative models are far below the experimentally achievable limits \cite{Nefkens&Price}.

  From the experimental point of view the only published result is that of
GAMS2000 experiment (1984) which has obtained the upper limit
$5\times10^{-4}$ at 90 \% C.L. \cite{PDG2002}. There is also a preliminary result from Crystall Ball collaboration (AGS/CB) at
Brookhaven which sets the limit at $1.8 \times 10 ^{-5}$ at 90 \% C.L 
\cite{Nefkens&Price}. 
Both the above experiments are of fixed target type, in which the $\eta$ meson is 
produced through the reaction $\pi ^{-}p \rightarrow \eta n$ with $P_{\pi
  ^{-}} = 30$ GeV/c for
 GAMS and 760 MeV/c for AGS/CB. \\

\vspace{0.2cm}

\section{Kloe detector}

The detector consists of a large cylindrical drift chamber, DC
\cite{K-DC}, surrounded by a lead-scintillating fiber sampling
calorimeter, EMC \cite{K-EMC},  both immersed in a solenoidal magnetic
 field of 0.52 T with the axis parallel to the beams.
The DC tracking volume extends from 28.5 to 190.5 cm
in radius and is 340 cm in length. For charged particles the
transverse momentum resolution is $\delta p_{T}/p_{T} \simeq
0.4\%$ and vertices are reconstructed with a spatial resolution of
$\sim$ 3 mm. 
The calorimeter is divided into a barrel and two endcaps
and covers 98$\%$ of the solid angle. 
Photon energies and
arrival times are measured with resolutions $\sigma_{E}/E =
0.057/{\small \sqrt{E \ ({\rm GeV})}}$  and $\sigma_{t} = 54 \ {\rm ps}
/{\small \sqrt{E \ ({\rm GeV})}} \oplus 50 \ {\rm ps}$  respectively. 
The photon entry points are determined with an accuracy 
 $\sigma_l \sim  1 \ {\rm cm} /{\small \sqrt{E \ ({\rm GeV})}}$ 
along the fibers, and $ \sim 1 $ cm in the transverse directions.
A photon is defined as a calorimeter cluster not associated to a charged 
particle, by requiring that the distance along the fibers
between the cluster centroid and the impact point of the nearest 
extrapolated track be greater than 3$\sigma_l$.
Two small calorimeters, QCAL \cite{K-QCAL}, made with lead and
scintillating tiles are wrapped around the low-beta quadrupoles to
complete the hermeticity.

The trigger \cite{K-TRIGGER} uses information from both the
calorimeter and the drift chamber. The EMC trigger requires
two local energy deposits above threshold 
($E > 50$ MeV in the barrel, $E > 150$ MeV in the endcaps). 
Recognition and rejection of cosmic-ray events is also performed at the
trigger level, checking for the presence of two energy deposits above 30
MeV in the outermost calorimeter planes.
The DC trigger is based on the multiplicity
and topology of the hits in the drift cells. 
The trigger has a large time spread  with respect to the beam crossing
time.
It is however synchronized with the machine radio frequency divided by four,
$T_{\rm sync}$ = 10.85 ns, with an accuracy of 50 ps.
During the period of data taking the bunch crossing period at 
DA$\Phi$NE was $T$ = 5.43 ns. 
The $T_0$ of the bunch crossing producing an event
is determined offline during the event reconstruction.


\section{Background}

The signal in this analysis is $\phi\to\gamma \eta(\to 3\gamma) \to
4\gamma$, the cross section corresponding to the PDG2002 upper limit is very low:  
$<$ 0.0215 nb (at KLOE the $\phi$ production cross section is about 3 $\mu$b).
So, background studies should 
cover all possible neutral processes at KLOE. 
The most important process which gives 4 photons in the final state is 
 $e^+e^-
\to \omega \gamma$ with $\omega \to \pi^0  \gamma$ but also important
are processes with less or more than 4 photons in final state which mimic 4
photons events because of cluster splitting or merging and accidental
clusters due to machine background.
The agreement with MC in the
background
description is very hard to obtain due to the above effects which 
are very difficult to reproduce; so
we have estimated the background directly from data and used the MC only to
evaluate the detection efficiency of the signal. \\

To this aim 
a generator of $\eta\to$$\gamma\gamma\gamma$ with pure phase space for the
decay dynamics 
has been used to produce 120,000
$\phi\to\gamma\eta(\eta \to 3\gamma)$ events.

\section{Event selection}

\vspace{0.3cm}

First a preselection is applied to the reconstructed events of the neutral
 radiative stream.
A  'recover splitting'  procedure (applied both
to data and to MC events) is implemented to reduce the number of 
split clusters; moreover it is required that  every cluster does not have an  association
with a DC track. Reconstructed data also include beam position  [8] and
 $\phi$ energy momentum
determination,
obtained run by run from the analysis of Bhabha events.
 MC includes cluster time offset correction, and simulation of software filters used
 to select events for the neutral radiative stream.

The following cuts are applied to the events:

\vspace{0.45cm}

  1) Ntime=4

\vspace{0.25cm}

Requires 4 photons with the correct time of flight $t$ : $|t-r/c| < 5
\sigma_t$ ; $r$ being the distance of the
cluster from I.P. and $\sigma_t$ being the calorimeter time resolution.

\vspace{0.45cm} 

  2) Ngood  = 4 

\vspace{0.25cm}

A good photon is defined in the following way: cluster energy $>30$MeV, \makebox{$|cos\theta|<0.93$}, where $\theta$ is the angle respect to the beam direction. 
Cluster energy cut helps to reject accidental clusters and split clusters, while
the polar angle cut 
excludes the blind region around the beam-pipe.

\vspace{0.45cm}

  3) Total energy of prompt clusters $>$ 800 MeV

\vspace{0.25cm}

 4) Total momentum $<$ 200MeV 

\vspace{0.25cm}

These last two cuts reject events from channels with more than 4 photons in
the final state which give background if some photons are
lost.

The events that pass the above preselection criteria are stored in reduced
files and to them are applied the following additional cuts: \\

 5)  $\theta_{\gamma\gamma}$$>$15$^{\circ}$

\vspace{0.25cm}

$\theta_{\gamma\gamma}$ is the minimum angle between two photons, to
further reject split clusters, which mainly come from 3$\gamma$ final
states.

\vspace{0.45cm}

  6) $E_{min} > 50$ MeV, $|cos\theta|$ $<$ 0.91.

\vspace{0.25cm}

 These cuts further reject 
accidental clusters and QED background.

\vspace{0.25cm}

\section{Kinematic fit}

At this point a kinematic fit procedure is applied to the four photons  to
improve the energy resolution. The input variables of
the fit are the following:
\begin{itemize}
\item X,Y,Z coordinates of the cluster; 
\item E energy of the cluster; 
\item t time of the cluster; 
\item X,Y,Z of the interaction vertex; 
\item E,Px,Py,Pz of the $\phi$ meson.
 
\end{itemize}
\vspace{0.25cm}

The fit is made according to the Lagrange's multipliers method, minimizing the following $\chi ^2$:
\begin{equation}
  \chi^2=\sum_{i}{\frac{(x_i-\mu_i)^2}{\sigma_i^2}}+\sum_{j}
  {\lambda_j F_j(\mu_k)}
\end{equation}
with 27 free parameters, while the $ F_j(\mu_k) $ represent the energy,
momentum and time constraints. \\

\begin{figure}
\begin{center}
\includegraphics[scale=0.45]{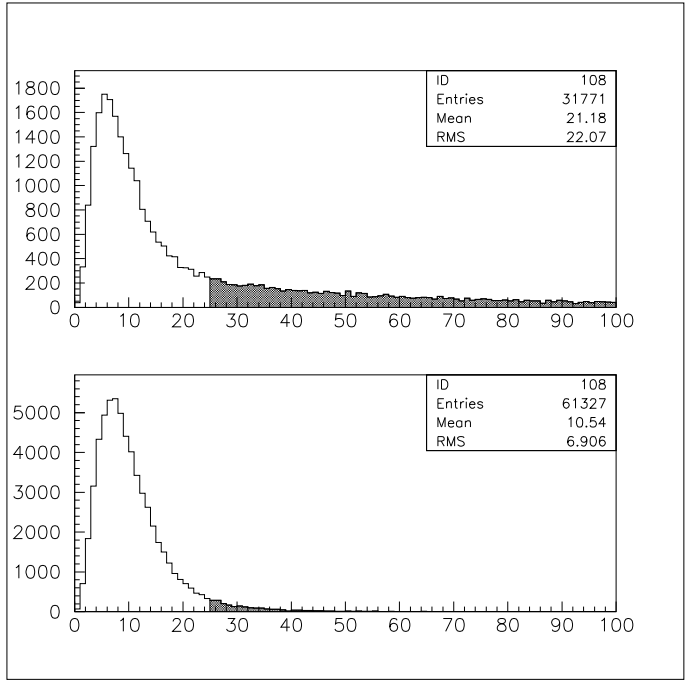}
\end{center}
\caption{$\chi ^2$ of kinematic fit distribution for: top DATA,  
bottom MC signal. Events with $\chi^2 > 25$ are rejected. } \label{fig:chi}
\end{figure}

\begin{figure}[hbtp]
\begin{center}
\includegraphics[scale=0.45]{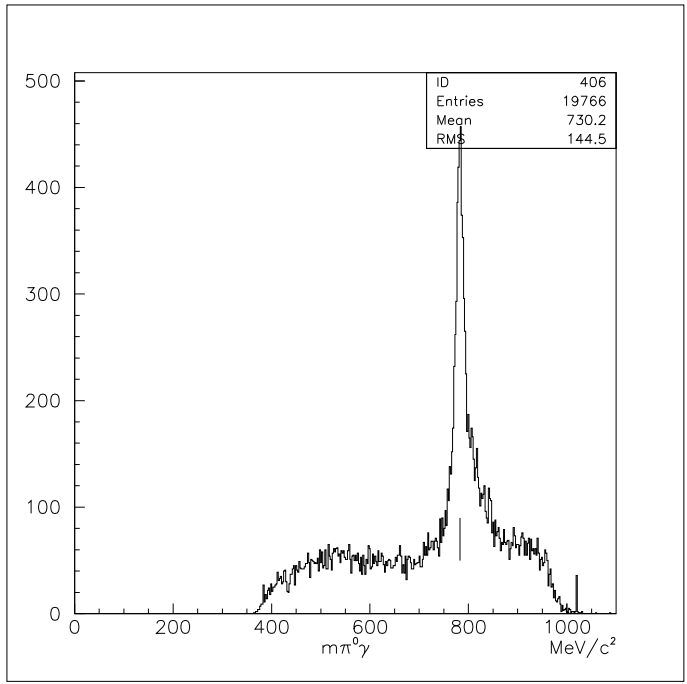}
\end{center}
\caption{ $m_{\pi ^0 \gamma}$  distribution, it shows a clear  peak at
 the $\omega$ mass.} \label{fig:mp0g}
\end{figure}

The $\chi ^2$ of the kinematic fit is used to reject the background
 using the cut $\chi ^2_{min} > 25$ (see Fig. \ref{fig:chi}).
At this point the main source of background is given by the channel $e^+e^- \to \omega ( \to \pi^0 \gamma ) \gamma $ where
a photon comes from ISR radiation.
This is evident by making a kinematic fit of the event in the $\pi^0 \gamma
\gamma$ hypothesis and looking to the plot of the 
$m_{\pi^0 \gamma}$ variable (choosing the most energetic one between the two photons
 not linked to the $\pi ^0$) (see Fig. \ref{fig:mp0g}).

\begin{figure}[hbtp]
\begin{center}
\includegraphics[scale=0.45]{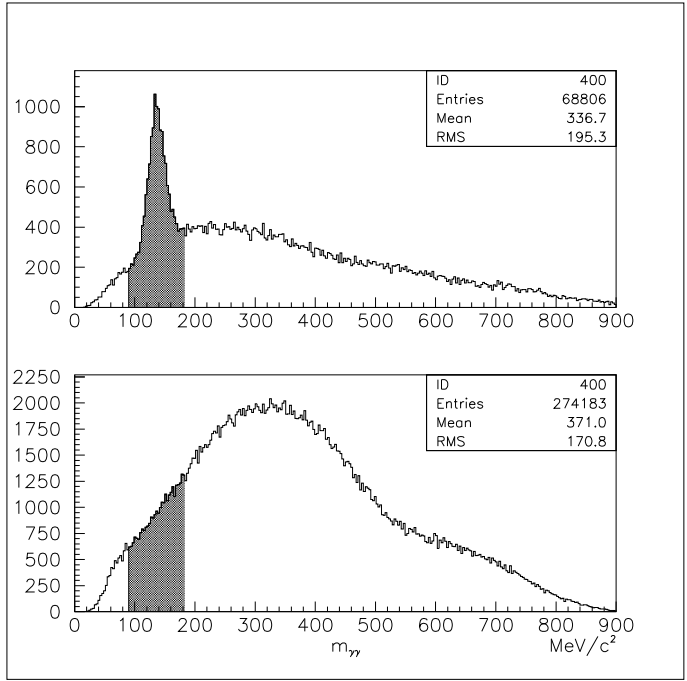}
\end{center}
\caption{$m_{\gamma \gamma}$ distributions. Top: 2001 DATA, bottom:
  MC signal. Events in the 90 - 180 MeV mass window are rejected.} \label{fig:mgg}
\end{figure}

Events with a $\pi^0$ in the final state represent a very large
fraction of background, they come
also from $\phi \to f_{0} \gamma \to \pi^0 \pi^0 \gamma$, from $\phi \to a_{0} \gamma \to \eta \pi^0 \gamma$ and $\phi \to \pi^0 \gamma$
final states. We reject these events containing a $\pi^0$ by cutting on
 the invariant mass $m _{\gamma \gamma}$ built from any couple of photons.
This variable is plotted in Fig.\ref{fig:mgg} both for 2001 data and MC signal. The cut chosen
is $90~ \textrm{MeV}/c^{2} <  m_{\gamma \gamma} < 180~ \textrm{MeV}/c^2$.

\section{Upper limit evaluation}

\begin{figure}
\includegraphics[width=0.5\textwidth]{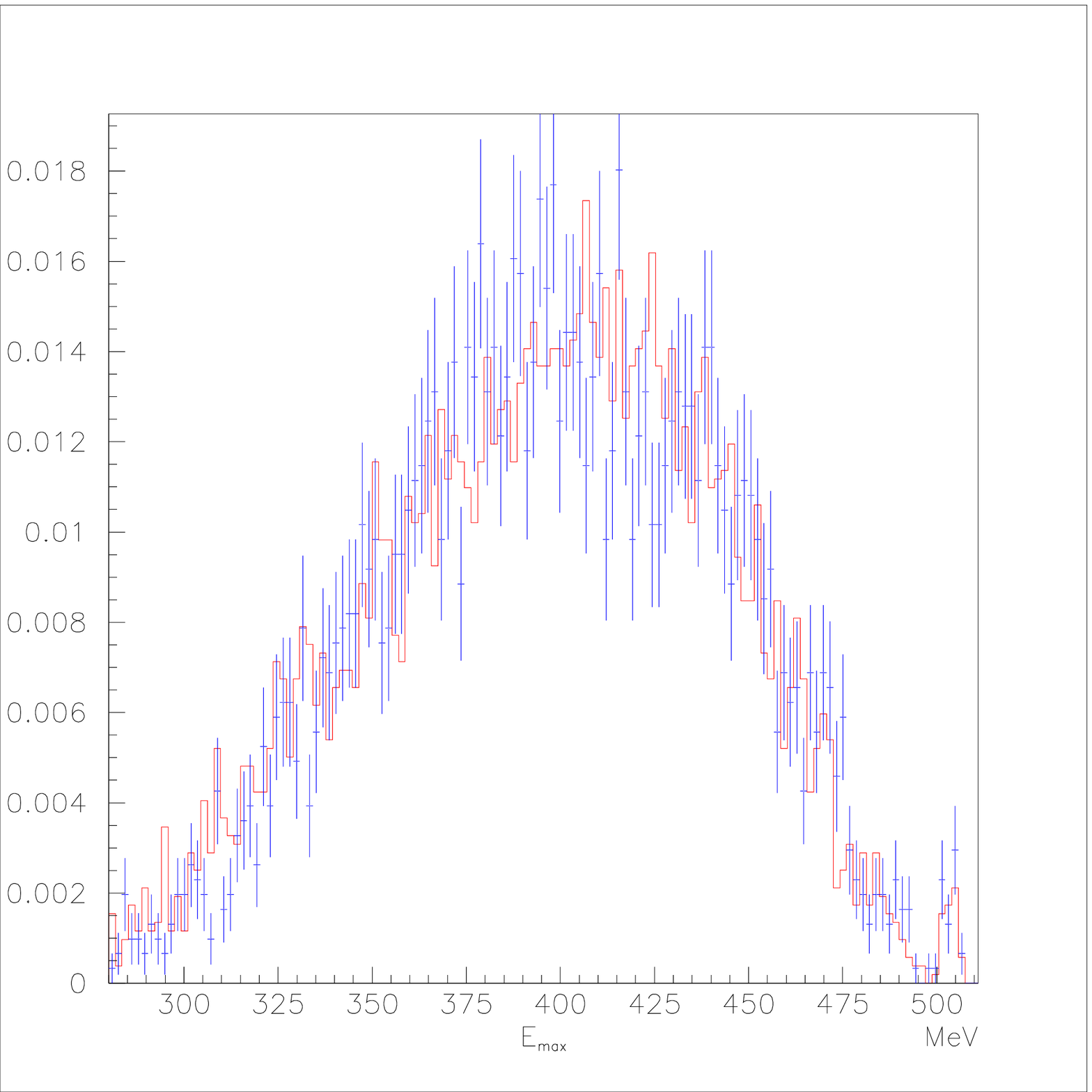}
\includegraphics[width=0.5\textwidth]{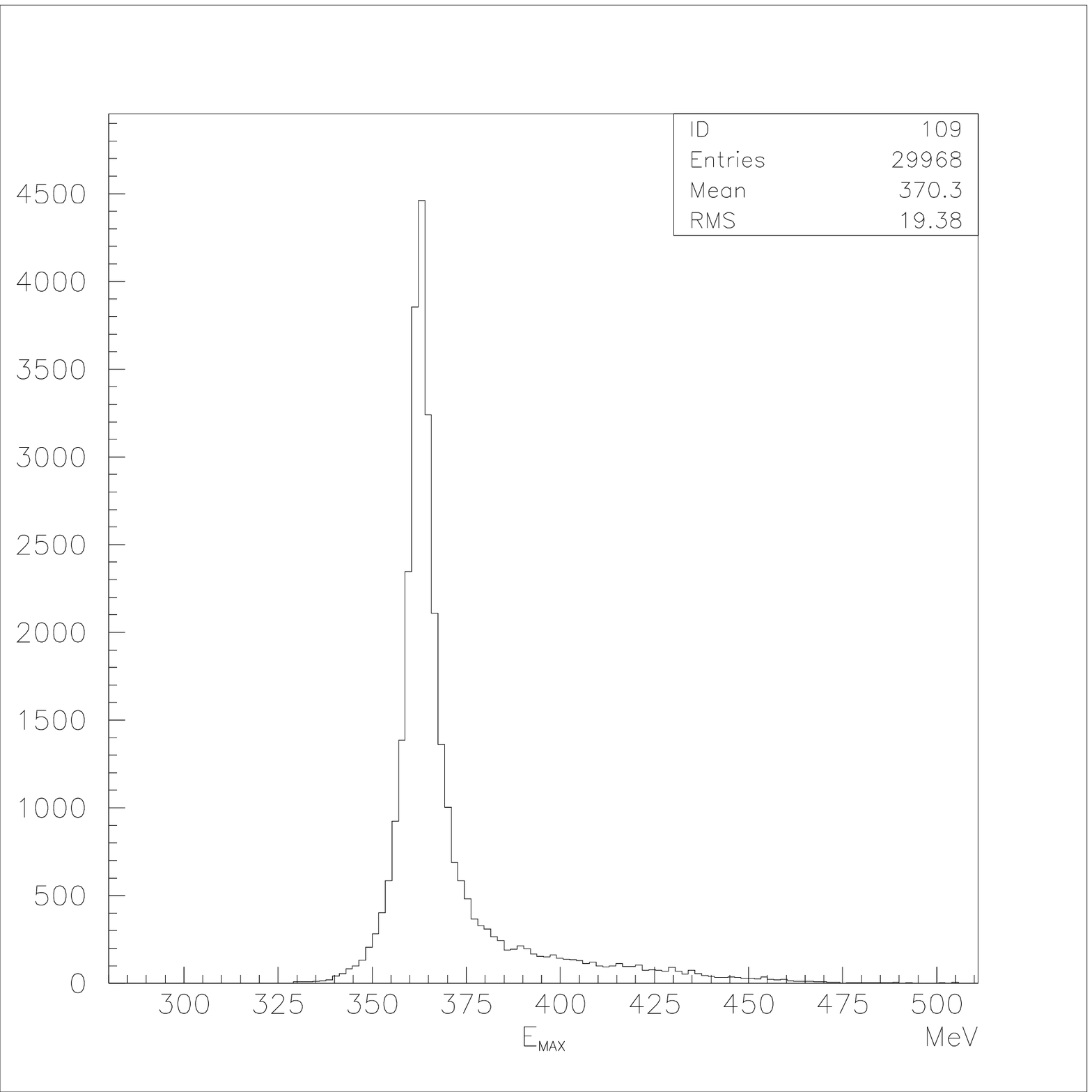}
\caption{Spectrum of the most energetic photon in the $\phi$ reference frame:
left: 2002 data (red histogram) and 2001 data (blue points); right: MC signal $\eta \to 3\gamma$} \label{fig:maxenergy}
\end{figure}

To search for the $\eta \to 3 \gamma$ events, we look for a peak in the
 distribution of the energy $E_{max}$ (evaluated in the 
$\phi$ reference frame) of
 the most energetic photon .  For the majority of the events this photon corresponds to the 
radiative one which has an expected energy of about 363 MeV.
  In Fig.\ref{fig:maxenergy} is shown 
the $E_{max}$ distribution for MC, 2001  and 2002 data.

As can be seen the two distributions for 2001
and 2002 overlap very well. A Kolmogorov test has been performed to check their compatibility, it gives a compatibility probability of 26\%,
 so we use the whole sample.
From the distribution is very clear that there isn't a narrow peak
 in  $E_{max}$, so we don't see any evidence of $\eta \to 3 \gamma$ events.

\begin{figure}[hbtp]
\begin{center}
\includegraphics[scale=0.45]{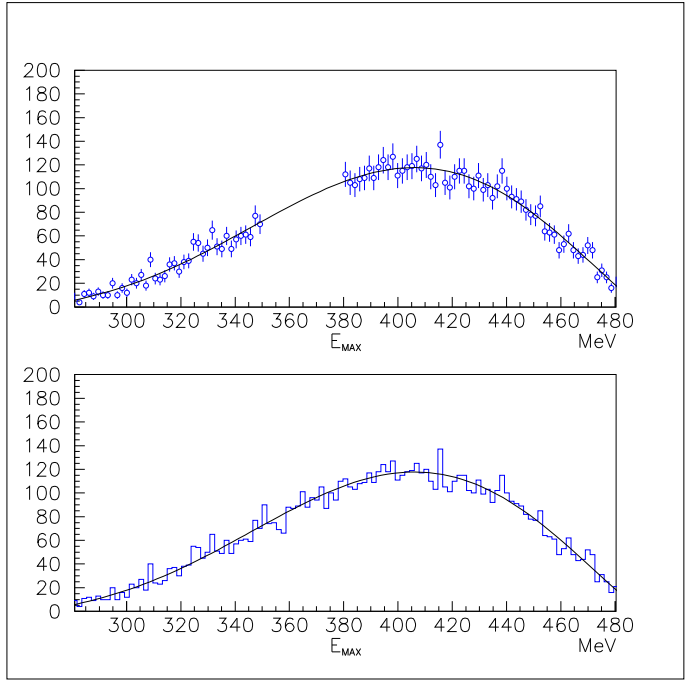}
\end{center}
\caption{Top:  $E_{max}$ distribution for 2001+2002 data. The interval
  used for background estimation
 is shown;
bottom: the fitted polynomial is overlapped to the whole range.} \label{fig:polfit}
\end{figure}

Then to evaluate the upper limit on the $\eta \to 3\gamma$ signal
using this distribution we choose as signal region (the region
where there is the main part of the signal)  the interval [350,380] MeV.
Then we fit the observed distribution of $E_{max}$  in the domain [280,350] MeV $\cup$ [380,480] MeV with a fifth degree polynomial, 
The fit is good ($\chi^2$/n.d.o.f = 77.9/92);
 (see Fig. \ref{fig:polfit}) and the fitted polynomial is used to estimate the number of the background events
in the signal region [350,380].

\begin{figure}
\begin{center}
\includegraphics[scale=0.6]{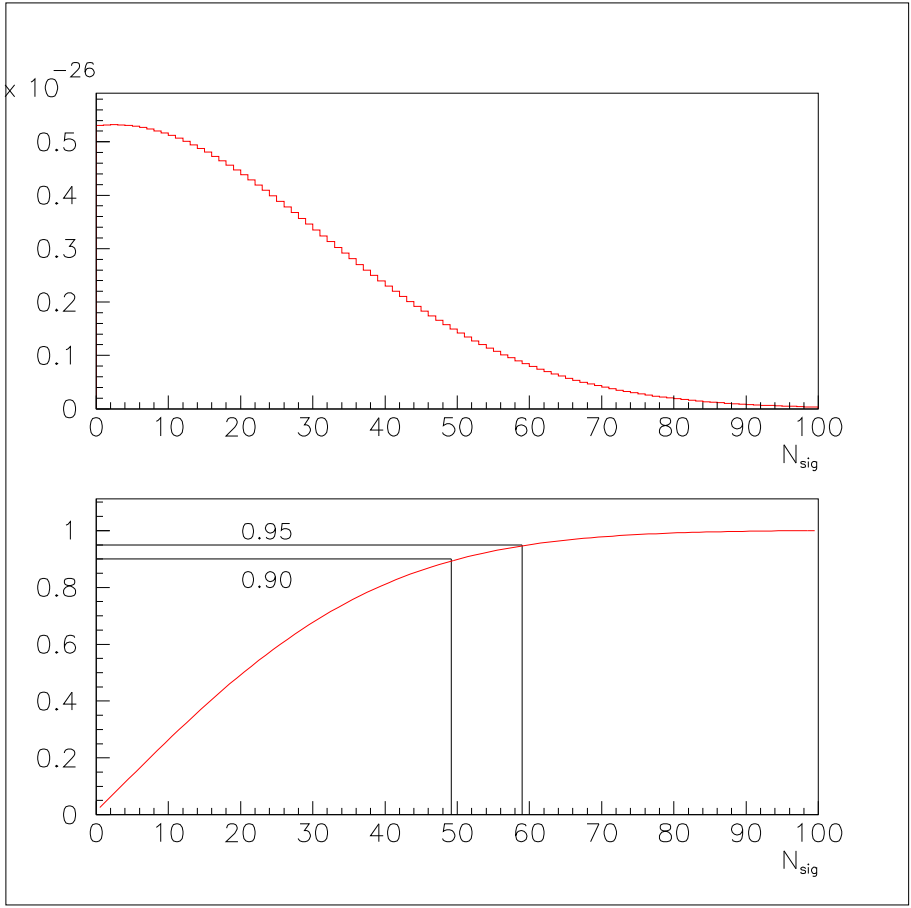}
\end{center}
\caption{Top: Likelihood as a function of $N_{sig}$; bottom: 
Integral function of the likelihood; the 
95 \% and 90 \% limits are shown.} \label{fig:liky}
\end{figure}

  In this range we assume to have both background and 
signal and build a likelihood function in this way:
\begin{center}
\[
Likelihood(N_{sig}) =
 \prod _{i = 1}^{N_{bin}}\frac{1}{\sqrt{2\pi N^{exp}_i}}\cdot exp - \left[\frac{(N_i - N^{exp}_i)^2}{2N^{exp}_i}\right]
\]
\end{center}
where $N_i$ is the number of observed events in the bin ``i'' (we use in
the signal region 15 bins of 1.75 MeV width), and $N^{exp}_i$ is the
number of the expected events given by:
\[
N^{exp}_i = a_{1}+a_2\cdot x_i+a_3\cdot x^2_i+a_4\cdot x^3_i+a_5\cdot x^4_i+a_6\cdot x^5_i+N_{sig}\cdot f_i
\]
where $x_i$ is the central value of the bin ``i'' and $f_i$ is the fraction
of signal events that fall in the bin ``i'' ($\sum_i f_i = 1$).
The $f_i$ are evaluated using the histogram of Fig. \ref{fig:maxenergy}(right).    
The likelihood function and its integral function are shown in Fig. \ref{fig:liky}.
We obtain the following values for the upper limit on $N_{sig}$:
\[
N_{sig} \le 59.0 \quad \textrm{at 95 \% C.L.} 
\]
\[
N_{sig} \le 49.2 \quad \textrm{at 90 \% C.L.} \\
\]

To convert the above results into the corresponding upper limits on the \makebox{Br($\eta \to 3\gamma$)}
we have evaluated the number of $\eta$ in the data sample through the study of the decay channel
$\eta \to 3 \pi^0$ by using an already developed analysis inside the experiment \cite{Miscetti and Giovannella}.
The efficiency of the selection evaluated from Monte Carlo is $\epsilon = 20.3$ \%. 
The result is:
\makebox{$N(\eta \to 3 \pi ^0) = 2431917$} and $\epsilon_{\eta \to 3\pi ^0}
= 0.4106 \pm 0.0016$.
 Then we evaluate the 
ratios of the two branching ratios:
\[
\frac{Br(\eta \to 3 \gamma)}{Br(\eta \to 3 \pi ^0)} \le
\frac{N_{Up}~\epsilon_{\eta \to 3 \pi ^0}}
{N_{\eta \to 3 \pi^0}~ \epsilon_{\eta \to 3 \gamma}}
\]

\[
\frac{Br(\eta \to 3\gamma)}{Br(\eta \to 3 \pi ^0)} \le 4.9 \cdot 10^{-5} \quad \textrm{95 \% C.L.}
\]

\[
\frac{Br(\eta \to 3\gamma)}{Br(\eta \to 3 \pi ^0)} \le 4.1 \cdot 10^{-5} \quad \textrm{90 \% C.L.}
\]

Using the PDG2002 value for $Br(\eta \to 3 \pi ^0) =  (32.51 \pm 0.29) \%$
we have:

\[
Br(\eta \to 3 \gamma) \le 1.6 \cdot 10 ^{-5} \quad \textrm{95 \% C.L.}
\]

\[
Br(\eta \to 3 \gamma) \le 1.3 \cdot 10 ^{-5} \quad \textrm{90 \% C.L.}
\]

\section{Systematic errors}

In this section we evaluate various systematic effects, especially those
which may arise from differences between the data and MC distributions used in the analysis.

\subsection{Systematics due to the $\chi ^2 cut$}

\begin{figure}[!hbtp]
\begin{center}
\includegraphics[scale=0.43]{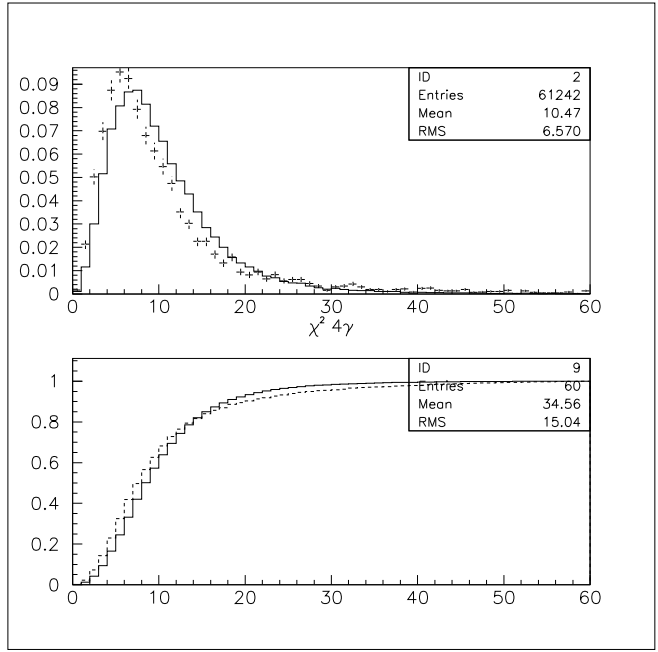}
\end{center}
\caption{Top: $\chi^2$ distribution for MC $\eta \to 3\gamma$, continuous
  line,
 and $\omega \gamma$ selected sample (see text), dashed line. Bottom: 
 integral of the above distributions.}
\label{fig:chiconf}
\end{figure}

We evaluate the systematics due to possible differences in the   $\chi
^2 $ distribution between data and MC by
comparing a data control sample with MC. The chosen control sample is 
the channel $e^+e^- \to \omega (\to \pi^0 \gamma) \to 4 \gamma$ that is the only channel with four photons that we have in 
our selection. Since the kinematic fit requires only the energy-momentum
conservation, 
we can compare its  $\chi ^2 $ distribution directly
to that of the channel $\eta \to 3 \gamma$ from MC.  
We select the $\omega \gamma$ candidates by requiring the mass range $128
< m_{\gamma \gamma} < 145$ $MeV/c^2$ for the  $\pi^0$ and
  $760 < m_{\pi^0 \gamma} < 815~ MeV/c^2$ for the $\omega$.
 In Fig. \ref{fig:chiconf} we have reported the $\chi^2$ distribution of
the kinematic fit, under energy-momentum conservation hypothesis, 
for the selected $\omega \gamma$ sample and for the MC $\eta \to 3\gamma$ sample. The upper
plot is the $\chi^2$ distribution, the lower plot is the fraction of events that survives to a given $\chi^2$ cut, normalized to the range shown in
Fig.  For $\chi^2_{cut} = 25$ the data-MC discrepancy is about 3\%. This is the systematic error that we assume for this cut.

\subsection{Correction for photon detection efficiency}

The MC doesn't simulate with full accuracy the photon detection efficiency, from the study done by \cite{Spadaro} 
we know, as a function of
the energy of the photon and the $cos(\theta)$, the ratio $w = \epsilon _{data}/\epsilon_{MC}$. 
For this reason, we have evaluated the quantity:
\[
Weight = \prod_{i = 1}^{4} w_i \quad \textrm{``i'' runs on the four photons.}
\]
for each event.
The sum of the weights gives  the effect of this discrepancy on the efficiency.
The results are:

\begin{center}
\begin{tabular}{|c|c|}
\hline \hline
unweighted events & 24376/120000 (20.3 \%) \\
\hline
weighted events & 24167/120000 (20.1 \%) \\
\hline
$\Delta \epsilon /\epsilon$ & 1.0 \% \\
\hline \hline
\end{tabular}
\end{center}

\subsection{Systematics due to signal or background shape and window choice }

\begin{figure}[!htbp]
\begin{center}
\includegraphics[scale=0.5]{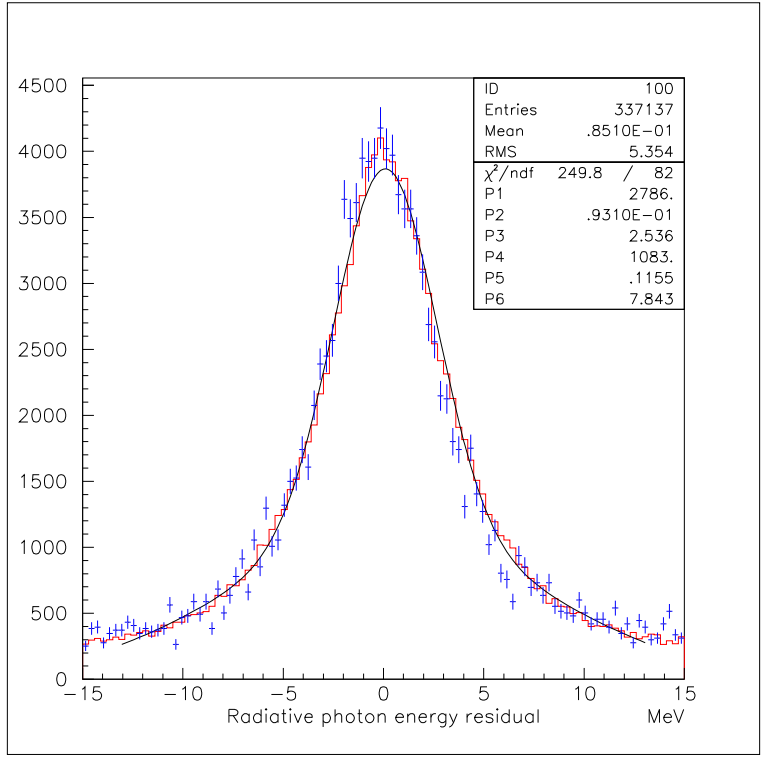}
\end{center}
\caption{$E_{\gamma} - E_{\gamma rad}$ distribution for  photons in the  $\phi \to \eta (\to \gamma
  \gamma) \gamma$ sample.
 Each event enters three times in the histogram, one for each photon. Both 
 MC (continuous line) and data
(points) are shown. $E_{\gamma rad}$ is the expected energy of the radiative
 photon = 362.7 MeV. The MC shape has been fitted with a double gaussian
 fit: 
$P1\cdot exp[-\frac{(x-P2)^2}{2\cdot P3^2}] +P4\cdot exp[-\frac{(x-P5)^2}{2\cdot P6^2}]$.} \label{fig:erad2g}
\end{figure}

To test if the kinematic fit introduces a bias in the photon energy that is different for data and MC,
we have analyzed a sample of $3\gamma$  events $(\phi \to \eta (\to \gamma
\gamma) \gamma)$
 and obtained from the kinematic fit  the updated energy for every photon. This energy is
plotted in Fig. \ref{fig:erad2g} both for data and MC. After having taken in account the wrong $\eta$ mass
that we have in MC generator ($m_{\eta} = 548.8$ MeV instead of 547.3 MeV) the two distribution overlap very well,
so we don't quote the systematics on this variable.

To take into account  other possible systematic effects in signal and
background shape 
we varied the width of the signal window and the order of the
polynomial used to fit the background shape.Then we repeated the fit of the background (background+signal)
excluding (including) the new energy window and by recalculating also the signal
efficiency: the maximum variation found
on the various upper limit estimates was about 35 $\%$.
\section{Conclusions}

  We have obtained the upper limit $ Br(\eta \to 3 \gamma) \le 1.6 \cdot 10
  ^{-5} \quad \textrm{at 95 \% C.L.}$ which is significantly better than the
  PDG existing one and also improves the unpublished Crystal Ball value. We
  estimate our systematic uncertainties to be at the $ 35 \% $ level. More
  data to be collected by Kloe in the next future will allow to further
  improve this result.

\section*{Acknowledgements}
We thank the DA$\Phi$NE team for their efforts in maintaining low background running 
conditions and their collaboration during all data-taking. 
We want to thank our technical staff: 
G.~F.~Fortugno for his dedicated work to ensure an efficient operation of 
the KLOE Computing Center; M.~Anelli for his continous support to the gas system and the safety of the detector; A.~Balla, M.~Gatta, G.~Corradi and G.~Papalino for the maintenance of the electronics; M.~Santoni, G.~Paoluzzi and R.~Rosellini for the general support to the detector; 
C.~Pinto (Bari), C.~Pinto (Lecce), C.~Piscitelli and A.~Rossi for their help during shutdown periods.
This work was supported in part by DOE grant DE-FG-02-97ER41027; by EURODAPHNE, contract FMRX-CT98-0169; by the German Federal Ministry of Education and Research (BMBF) contract 06-KA-957; by Graduiertenkolleg `H.E. Phys. and Part. Astrophys.' of Deutsche Forschungsgemeinschaft, Contract No. GK 742; by INTAS, contracts 96-624, 99-37; and by TARI, contract HPRI-CT-1999-00088.

\end{document}